\documentclass[a4paper]{report}
\usepackage[utf8]{inputenc}
\usepackage[T1]{fontenc}
\usepackage{RJournal}
\usepackage{amsmath,amssymb,array}
\usepackage{booktabs}

\providecommand{\tightlist}{%
  \setlength{\itemsep}{0pt}\setlength{\parskip}{0pt}}

\newlength{\cslhangindent}
\newlength{\csllabelwidth}
\newlength{\cslentryspacingunit} 
  {}%
  {\par}
\newenvironment{CSLReferences}[2] 
 {
  \setlength{\parindent}{0pt}
  \ifodd #1
  \let\oldpar\par
  \def\par{\hangindent=\cslhangindent\oldpar}
  \fi
  \setlength{\parskip}{#2\cslentryspacingunit}
 }%
 {}
\usepackage{calc}

\usepackage{longtable}
\definecolor{link}{RGB}{0,158,115}
\begin{document}

\begin{article}

\title{Coloring in R's Blind Spot}
\author{by Achim Zeileis and Paul Murrell}

\maketitle

\abstract{%
Prior to version 4.0.0 R had a poor default color palette (using highly saturated red, green, blue, etc.) and provided very few alternative palettes, most of which also had poor perceptual properties (like the infamous rainbow palette). Starting with version 4.0.0 R gained a new and much improved default palette and, in addition, a selection of more than 100 well-established palettes are now available via the functions \texttt{palette.colors()} and \texttt{hcl.colors()}. The former provides a range of popular qualitative palettes for categorical data while the latter closely approximates many popular sequential and diverging palettes by systematically varying the perceptual hue, chroma, luminance (HCL) properties in the palette. This paper provides an overview of these new color functions and the palettes they provide along with advice about which palettes are appropriate for specific tasks, especially with regard to making them accessible to viewers with color vision deficiencies.
}

\hypertarget{introduction}{%
\section{Introduction}\label{introduction}}

\begin{figure}[h!]

{\centering \includegraphics[width=1\linewidth]{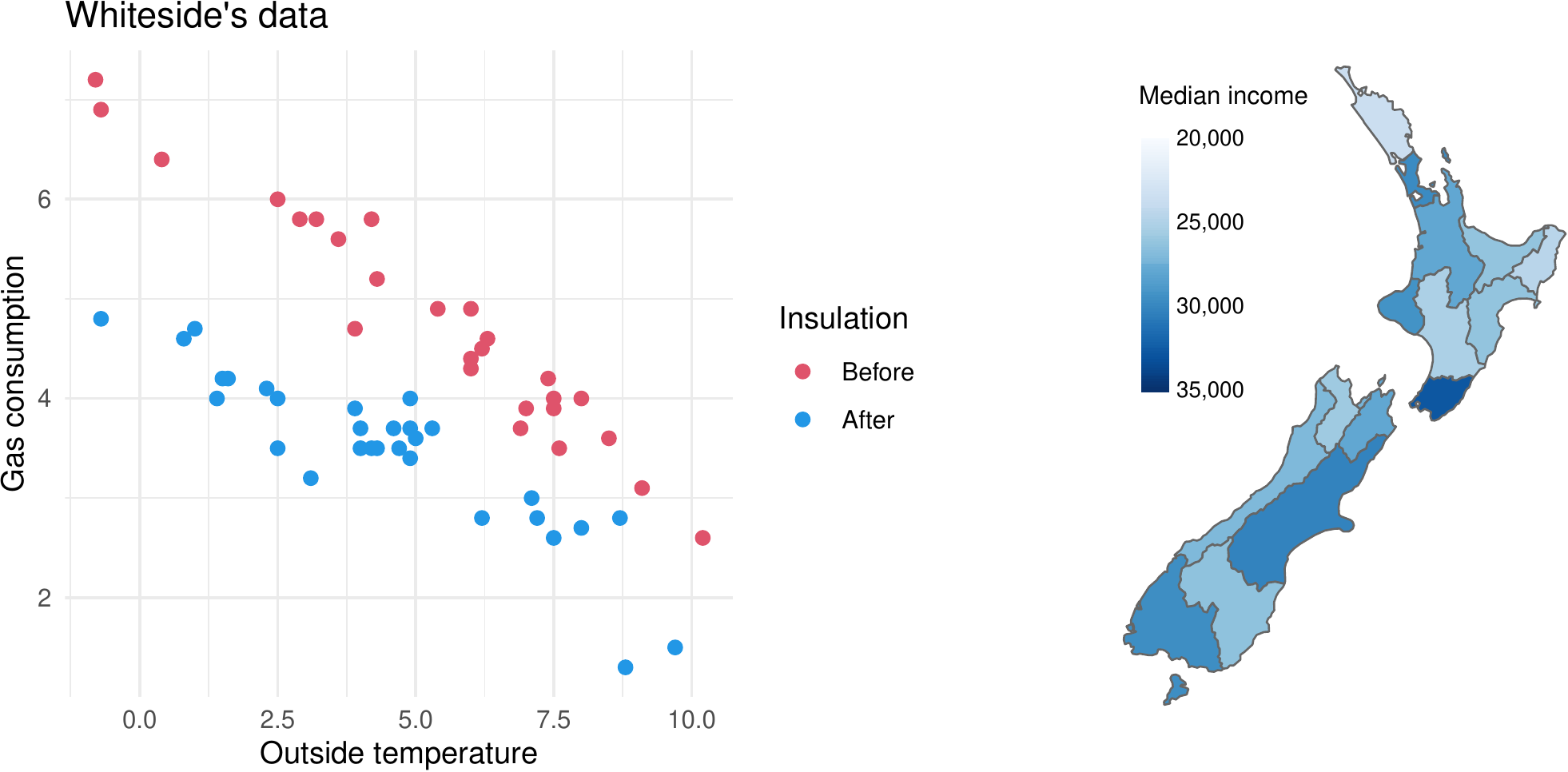} 

}

\caption{Typical usage of color for coding qualitative/categorical information (left) and quantitative/continuous information (right). Left: Scatter plot of weekly gas consumption by outside temperature before and after installing a house insulation. Right: Choropleth map of median income in the 16 regions of New Zealand in 2018.}\label{fig:colorcatcont}
\end{figure}

Color can be a very effective way to distinguish between different groups
within a data visualization. Color is a ``preattentive'' visual feature,
meaning that groups are identified rapidly and without conscious effort
(Ware 2012). For example, it is trivially easy to identify the
two groups of points in the scatterplot in Figure~\ref{fig:colorcatcont}.

Employing color to represent values on a continuous numeric scale will be
less successful (Cleveland and McGill 1984), but color can
still be useful to convey additional variables when more effective
visual features, such as location, have already been used. For example,
color might be used to fill in different regions on a map, as
demonstrated in the right hand plot of Figure~\ref{fig:colorcatcont}.

R provides several ways to specify a color: by name (e.g., \texttt{"red"});
by hexadecimal RGB code (e.g., \texttt{"\#FF0000"}); or by integer (e.g., \texttt{2}).
When we specify an integer, that provides an index into a
default set of colors; the color \texttt{2}
means the second color in the default set of colors.

However, a more important task than specifying one particular color
is the task of specifying a set of colors to use in combination
with each other.
For example, in the left panel of Figure~\ref{fig:colorcatcont},
we need two colors
that are very easily perceived as different from each other.
In the right panel of Figure~\ref{fig:colorcatcont},
we require a set of colors
that appear to change monotonically, e.g., from darker to lighter.

We call this the problem of selecting a good \emph{palette} of colors.
What we need to generate is a vector of R colors, e.g.,
\texttt{c("red",\ "blue")}, \texttt{c("\#FF0000",\ "\#0000FF")}, or \texttt{c(2,\ 4)}.

\hypertarget{a-brief-history-of-r-palettes}{%
\section{A brief history of R palettes}\label{a-brief-history-of-r-palettes}}

Early versions of R provided very few functions for choosing colors from readily available palettes and the palettes
that were provided, although standard at the time they were implemented,
have meanwhile
been widely recognized as being rather poor.

\begin{figure}[ht!]

{\centering \includegraphics[width=1\linewidth]{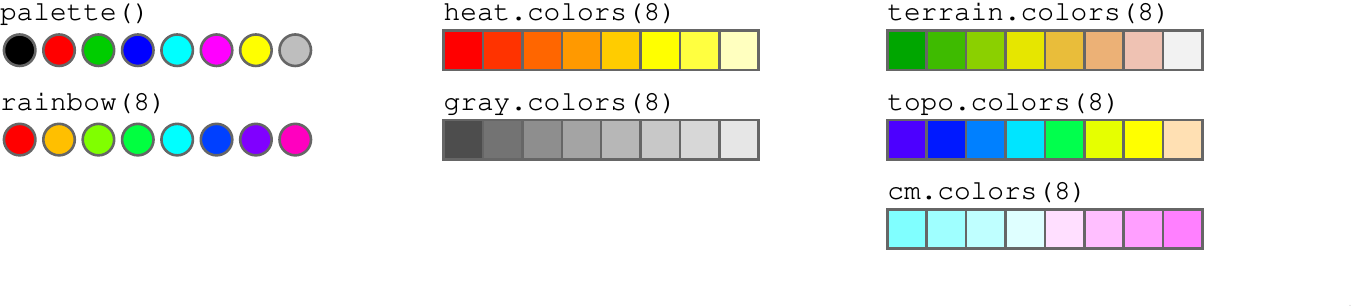} 

}

\caption{Old base R palettes.  At top left is the old default palette (prior to version 4.0.0), consisting largely of highly saturated primary colors or combinations thereof.  Below that is the rainbow palette of different highly saturated hues.  The middle column shows the old sequential palettes, with heat colors again being highly saturated.  The last column shows an old diverging palette plus two palettes motivated by shadings of geographic maps.}\label{fig:oldPalettes}
\end{figure}

The \texttt{palette()} function generates a vector of eight colors.
These provide the default set of colors that an integer color specification selects from
and can be used for coding \emph{categorical} information.
The output below shows what R used to produce prior to version 4.0.0, along with a
\emph{swatch} of color circles.

\begin{verbatim}
palette()
\end{verbatim}

\begin{verbatim}
#> [1] "black"   "red"     "green3"  "blue"    "cyan"    "magenta" "yellow" 
#> [8] "gray"
\end{verbatim}

\includegraphics[width=1\linewidth]{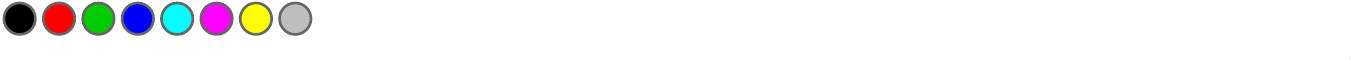}

Figure~\ref{fig:oldPalettes}
depicts this old default \texttt{palette()} (top-left) along
with other old base R palettes using swatches of circles or
rectangles that are filled with the
corresponding colors. The other palette functions all take an argument \texttt{n} to generate
that number of colors (possibly along with further arguments that allow for
certain customizations):

\begin{itemize}
\tightlist
\item
  \texttt{heat.colors()}, \texttt{terrain.colors()}, \texttt{topo.colors()}, and \texttt{gray.colors()}
  can be used as \emph{sequential} palettes for ordered or numeric information.
\item
  \texttt{cm.colors()} can be used as a \emph{diverging} palette for values that are
  distributed around a ``neutral'' value, such as zero.
\item
  \texttt{rainbow()} implements the infamous rainbow (or ``jet'') palette
  that was widely used (possibly with restrictions of the hue range)
  for all types of variables: \emph{categorical}, \emph{sequential}, and \emph{diverging}.
\end{itemize}

All of these palettes -- except \texttt{gray.colors()} -- have poor perceptual properties.
The colors are highly saturated, which can be distracting and overly
stimulating, and the colors are unbalanced with respect to chroma and luminance,
which means that they have unequal visual impact
(Lonsdale and Lonsdale 2019; Bartram, Patra, and Stone 2017; Etchebehere and Fedorovskaya 2017).
In addition, the palettes do not perform well for viewers with some form
of colorblindness (nearly 10\% of the male population).
Most of the palettes also use sequences
of hues obtained in the RGB (red-green-blue) space or simple derivations thereof like HSV (hue-saturation-value) or HLS (hue-lightness-saturation),
which leads to clustering of colors at the red, green, and blue primaries.

Although these limitations have been well known for some time, no changes
were made to these palettes provided by the core R graphics system for
a number of years. There were various reasons for this including the following:

\begin{itemize}
\item
  In R version 2.1.0, Thomas Lumley added the \texttt{colorRampPalette()} function.
  This made it easier to generate a palette, though the user is still required
  to select, for example, start and end colors from which
  a palette of colors can then be interpolated.
\item
  Better palettes became available via packages on CRAN
  (Comprehensive R Archive Network)
  starting with \CRANpkg{RColorBrewer} (Neuwirth 2022), later
  \CRANpkg{colorspace} (Ihaka 2003; Zeileis, Hornik, and Murrell 2009), and
  more recently
  \CRANpkg{viridis} (Garnier 2022), \CRANpkg{rcartocolor} (Nowosad 2019),
  \CRANpkg{scico} (Pedersen and Crameri 2022), and
  \CRANpkg{cols4all} (Tennekes 2023),
  among many others.
\item
  Higher-level graphics systems
  like \CRANpkg{ggplot2} (Wickham 2016) and \CRANpkg{lattice} (Sarkar 2008)
  developed their own color themes.
\end{itemize}

\hypertarget{a-new-set-of-r-palettes}{%
\section{A new set of R palettes}\label{a-new-set-of-r-palettes}}

On the road to R version 4.0.0 an attempt was made to
address the limited and deficient set of palettes in base R
and to add a range of modern color palettes.
In particular, \texttt{palette()} has a new improved default color palette,
\texttt{palette.colors()} provides further well-established qualitative palettes (Zeileis et al. 2019), and
\texttt{hcl.colors()} provides a wide range of qualitative, sequential, and diverging palettes obtained by a standardized approach in
the so-called HCL (hue-chroma-luminance) space (Wikipedia 2023); see Zeileis and Murrell (2019) and Zeileis et al. (2020).

\hypertarget{a-new-default-color-palette}{%
\subsection{\texorpdfstring{A new default color \texttt{palette()}}{A new default color palette()}}\label{a-new-default-color-palette}}

The default color palette in R -- the default set of colors that can
be specified by integer index -- has been replaced. The new palette
follows the same basic hues as the old default palette, but
the palette is less saturated overall and
reduces the size of changes in chroma and luminance across the palette.
This produces a calmer and less distracting palette with a more
even visual impact.
An attempt has also been made to improve the discriminability
of the colors in the default palette for colorblind viewers.
The output (and swatches) below show what R produces from version 4.0.0
onwards.

\begin{verbatim}
palette()
\end{verbatim}

\begin{verbatim}
#> [1] "black"   "#DF536B" "#61D04F" "#2297E6" "#28E2E5" "#CD0BBC" "#F5C710"
#> [8] "gray62"
\end{verbatim}

\includegraphics[width=1\linewidth]{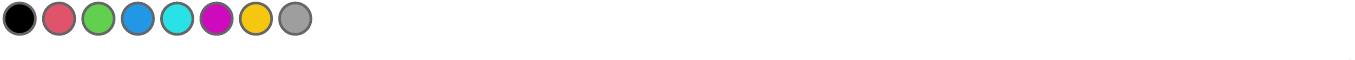}

\hypertarget{the-palette.colors-function}{%
\subsection{\texorpdfstring{The \texttt{palette.colors()} function}{The palette.colors() function}}\label{the-palette.colors-function}}

\begin{figure}[ht!]

{\centering \includegraphics[width=1\linewidth]{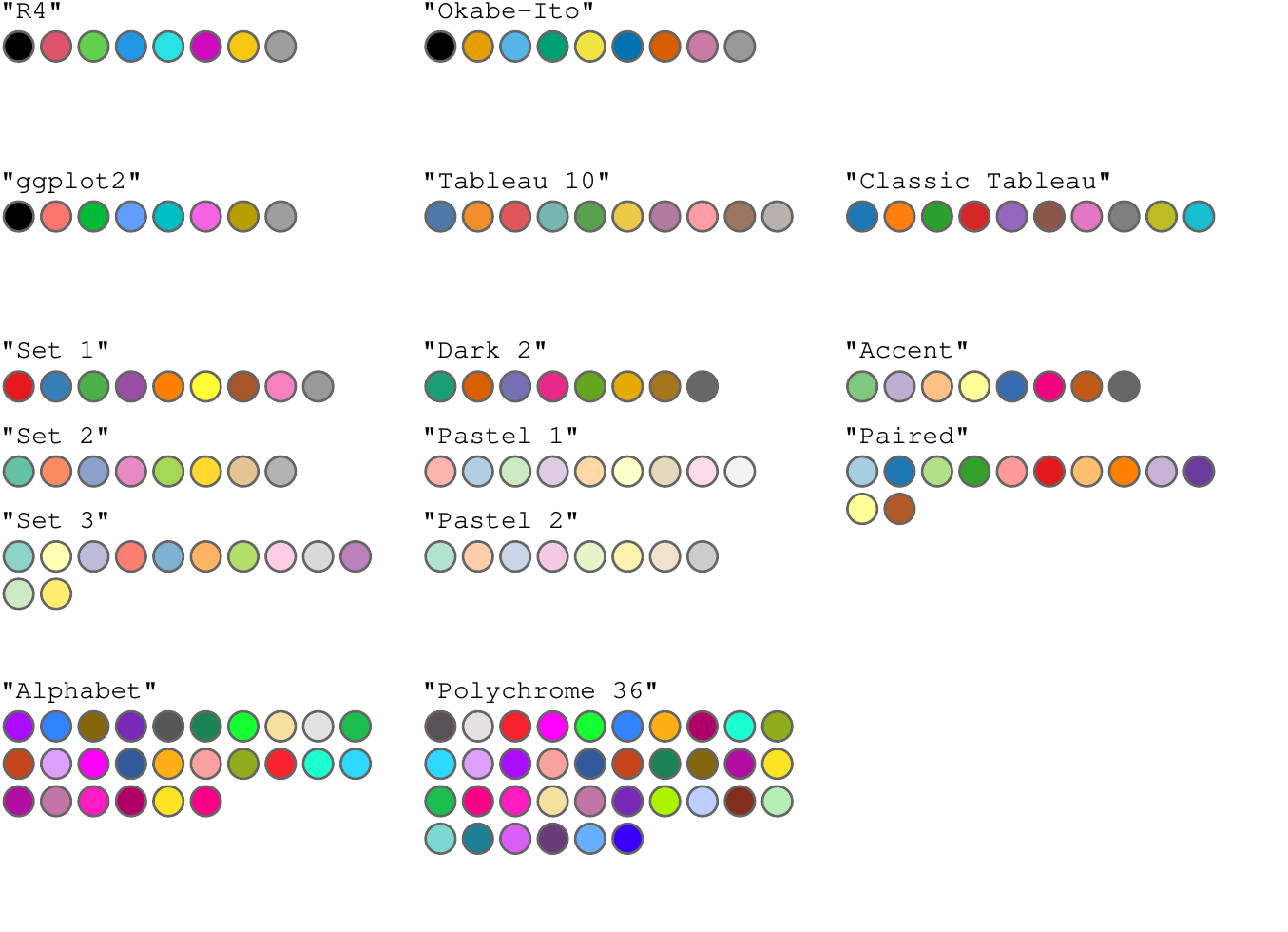} 

}

\caption{New qualitative palettes in base R available from the \code{palette.colors()} function.  The label above each swatch shows the argument to provide to \code{palette.colors()} to produce the set of colors.  The palette at top-left is the new default that is also produced by \code{palette()}.  The \code{"Okabe-Ito"} palette is the default that is produced by \code{palette.colors()} (with no arguments).}\label{fig:newPalettes}
\end{figure}

The \texttt{palette.colors()} function, new in R 4.0.0, provides a way to
access several other predefined palettes
(see also Figure~\ref{fig:newPalettes}).
All of these are \emph{qualitative palettes} so they are appropriate for
encoding qualitative (categorical) variables. In other words,
these palettes are appropriate for differentiating between groups.
By default \texttt{palette.colors()} returns the
\texttt{"Okabe-Ito"} (Okabe and Ito 2008) palette.
This palette
was designed to be very robust under color vision deficiencies, so
the different colors in this palette should be easily distinguishable
for all viewers.

\begin{verbatim}
palette.colors()
\end{verbatim}

\begin{verbatim}
#>         black        orange       skyblue   bluishgreen        yellow 
#>     "#000000"     "#E69F00"     "#56B4E9"     "#009E73"     "#F0E442" 
#>          blue    vermillion reddishpurple          gray 
#>     "#0072B2"     "#D55E00"     "#CC79A7"     "#999999"
\end{verbatim}

\includegraphics[width=1\linewidth]{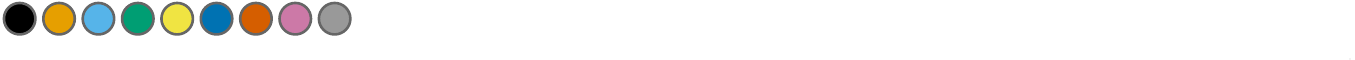}

The first argument to \texttt{palette.colors()} is a number of colors.
Each palette has a fixed number of colors, but we can ask for fewer or,
with \texttt{recycle\ =\ TRUE}, we can get more colors by recycling.
For example, the following code just requests the first four colors
from the \texttt{"Okabe-Ito"} palette.

\begin{verbatim}
palette.colors(4)
\end{verbatim}

\begin{verbatim}
#>       black      orange     skyblue bluishgreen 
#>   "#000000"   "#E69F00"   "#56B4E9"   "#009E73"
\end{verbatim}

\includegraphics[width=1\linewidth]{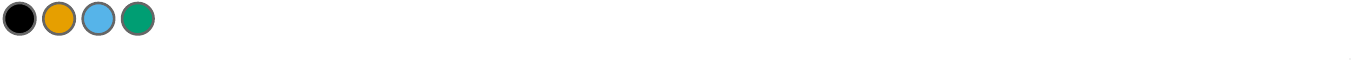}

The following code requests ten colors from the \texttt{"Okabe-Ito"} palette.
That palette only contains nine colors, but because \texttt{recycle\ =\ TRUE},
a tenth color is provided by recycling the first color (black) from the palette.

\begin{verbatim}
palette.colors(10, recycle = TRUE)
\end{verbatim}

\begin{verbatim}
#>         black        orange       skyblue   bluishgreen        yellow 
#>     "#000000"     "#E69F00"     "#56B4E9"     "#009E73"     "#F0E442" 
#>          blue    vermillion reddishpurple          gray         black 
#>     "#0072B2"     "#D55E00"     "#CC79A7"     "#999999"     "#000000"
\end{verbatim}

\includegraphics[width=1\linewidth]{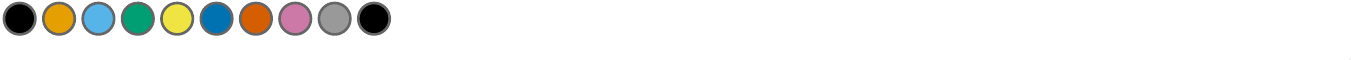}

The second argument to \texttt{palette.colors()} is the palette to select colors
from. For example, the following code requests the first four colors
from the \texttt{"R4"} palette (the new default in \texttt{palette()}).

\begin{verbatim}
palette.colors(4, palette = "R4")
\end{verbatim}

\begin{verbatim}
#> [1] "#000000" "#DF536B" "#61D04F" "#2297E6"
\end{verbatim}

\includegraphics[width=1\linewidth]{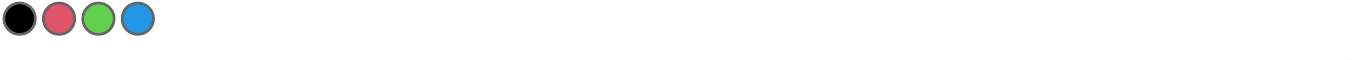}

\hypertarget{the-hcl.colors-function}{%
\subsection{\texorpdfstring{The \texttt{hcl.colors()} function}{The hcl.colors() function}}\label{the-hcl.colors-function}}

The \texttt{hcl.colors()} function was added in R 3.6.0, with the range
of supported palettes slowly expanded over time.
This function provides access to another range of palettes,
including sequential and diverging palettes
for representing continuous variables.
As with \texttt{palette.colors()}, the first
argument is a number of colors to generate and the second
specifies a palette to generate colors from.
The \texttt{hcl.pals()} function provides a full list of the
available palette names that we can choose from.

\begin{verbatim}
hcl.colors(8, palette = "Blues 3")
\end{verbatim}

\begin{verbatim}
#> [1] "#00366C" "#005893" "#007BC0" "#5E9BD8" "#91BAEB" "#BAD5FA" "#DDECFF"
#> [8] "#F9F9F9"
\end{verbatim}

\includegraphics[width=1\linewidth]{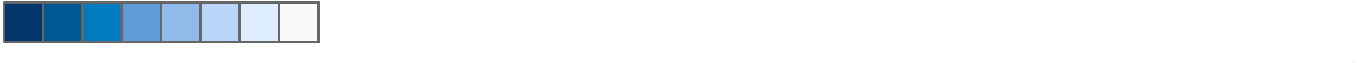}

\begin{figure}[ht!]

{\centering \includegraphics[width=0.49\linewidth]{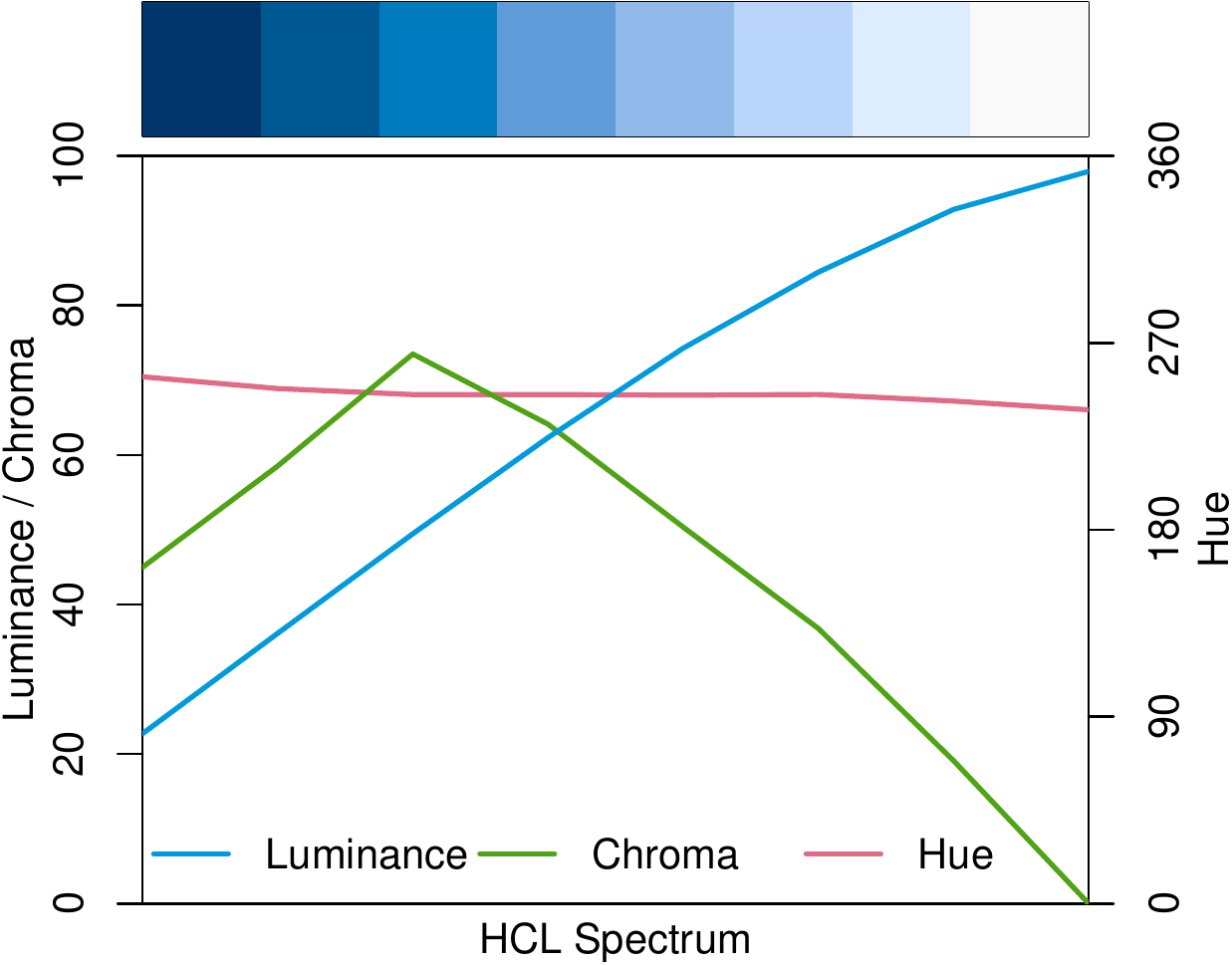} 

}

\caption{Hue, chroma, and luminance paths for the \code{"Blues 3"} palette. This plot is created by the \code{colorspace::specplot()} function.  We can see that hue is held constant in this palette, while luminance increases monotonically and chroma peaks towards the middle of the palette.}\label{fig:blues3hcl}
\end{figure}

One difference with \texttt{hcl.colors()} is that the palette we are
selecting colors from is \emph{not} a fixed set of colors. Instead,
the palettes in \texttt{hcl.colors()} are a path within HCL colorspace. For each
dimension -- hue, chroma, and luminance -- a palette can have a constant
value, a monotonic trajectory, or a triangular trajectory. For
example, the trajectories for the \texttt{"Blues\ 3"} palette are shown in
Figure~\ref{fig:blues3hcl}. The palette is (almost) constant in the hue
dimension yielding different shades of (almost) the same blue.
The palette is monotonically increasing in the luminance dimension, so
the blues vary from very dark to very light. Finally, the palette has a
triangular trajectory in the chroma dimension, so the blues are more
colorful towards the middle of the palette.

\begin{figure}[ht!]

{\centering \includegraphics[width=0.49\linewidth]{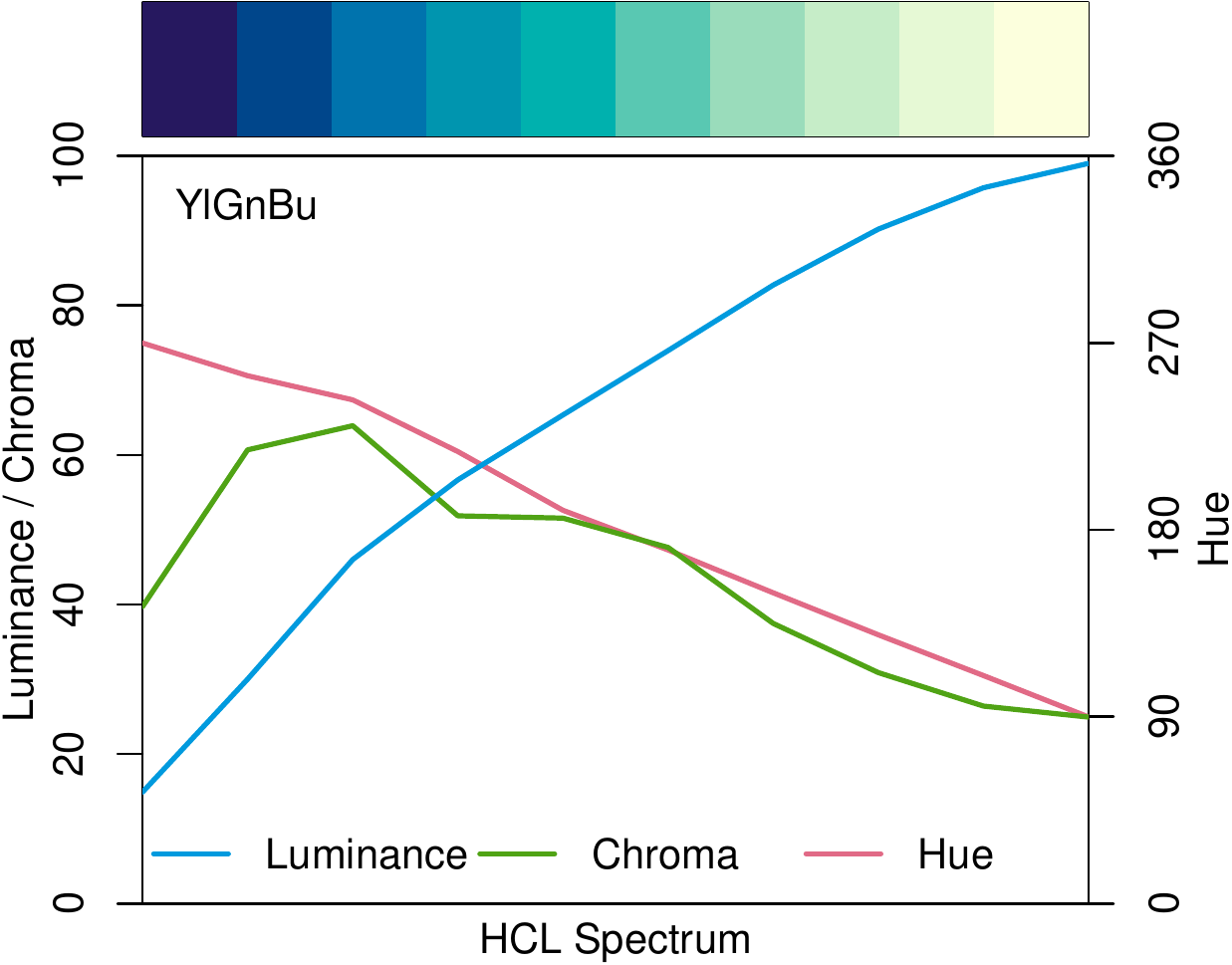} \includegraphics[width=0.49\linewidth]{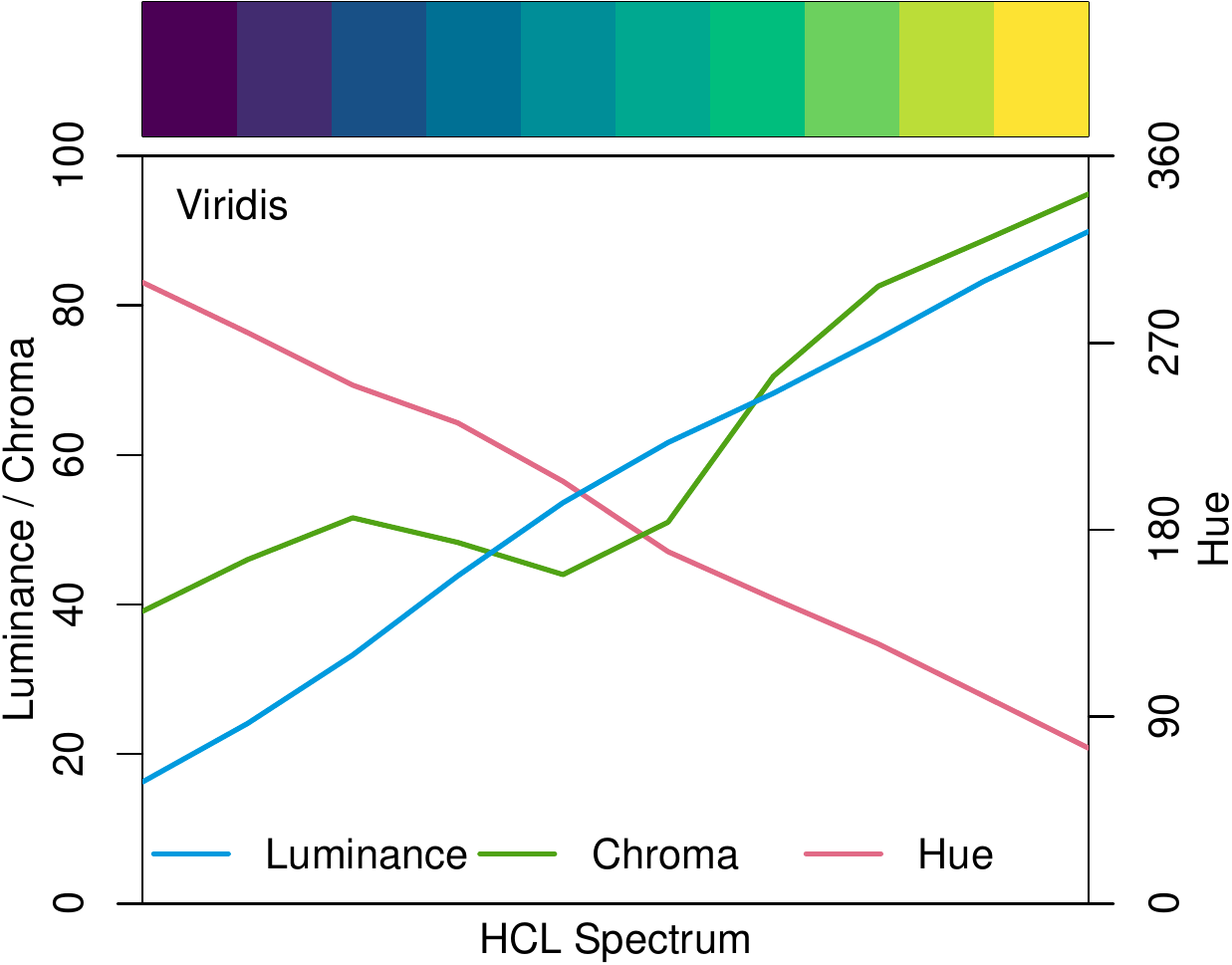} 

}

\caption{Hue, chroma, and luminance paths for the \code{"YlGnBu"} (left) and \code{"Viridis"} (right) palettes. These plots are created by the `colorspace::specplot()` function. For \code{"YlGnBu"} we can see that hue changes from blue to yellow, luminance increases monotonically, and chroma has a small peak in the blue range and then decreases with luminance. \code{"Viridis"}, on the other hand, has almost the same trajectory for both hue and luminance, but chroma increases for the light colors.}\label{fig:ylgnbu-viridis}
\end{figure}

\begin{figure}[ht!]

{\centering \includegraphics[width=0.49\linewidth]{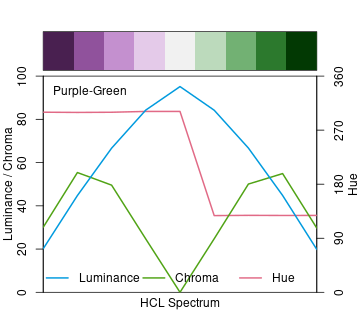} \includegraphics[width=0.49\linewidth]{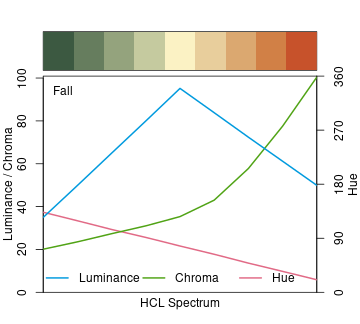} 

}

\caption{Hue, chroma, and luminance paths for the \code{"Purple-Green"} (left) and \code{"Fall"} (right) palettes. The plots are created by the \code{colorspace::specplot()} function. We can see that the \code{"Purple-Green"} palette is "balanced" with luminance and chroma varying symmetrically about the central neutral color for both hues. In contrast, the \code{"Fall"} palette is "unbalanced" with the left arm of the palette having somewhat darker colors with far less chroma than the right arm. Hue changes gradually from green through yellow to red, yielding a warmer palette compared to \code{"Purple-Green"}.}\label{fig:purplegreen-fall}
\end{figure}

Because the palettes from \texttt{hcl.colors()} are based on a continuous
path in HCL space, we can select as many colors as we like.
For example,
the following code generates five colors from the multi-hue sequential
palette \texttt{"YlGnBu"} (see also Figure~\ref{fig:ylgnbu-viridis})
and nine colors from the diverging palette \texttt{"Purple-Green"}
(see also Figure~\ref{fig:purplegreen-fall}).

\begin{verbatim}
hcl.colors(5, palette = "YlGnBu")
\end{verbatim}

\begin{verbatim}
#> [1] "#26185F" "#007EB3" "#18BDB0" "#BCE9C5" "#FCFFDD"
\end{verbatim}

\includegraphics[width=1\linewidth]{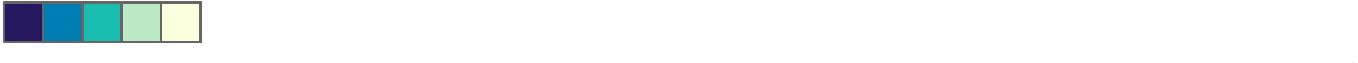}

\begin{verbatim}
hcl.colors(9, palette = "Purple-Green")
\end{verbatim}

\begin{verbatim}
#> [1] "#492050" "#90529C" "#C490CF" "#E4CAE9" "#F1F1F1" "#BCDABC" "#72B173"
#> [8] "#2C792D" "#023903"
\end{verbatim}

\includegraphics[width=1\linewidth]{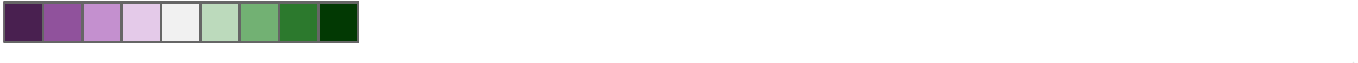}

\hypertarget{illustrations}{%
\subsection{Illustrations}\label{illustrations}}

\begin{figure}[h!]

{\centering \includegraphics[width=1\linewidth]{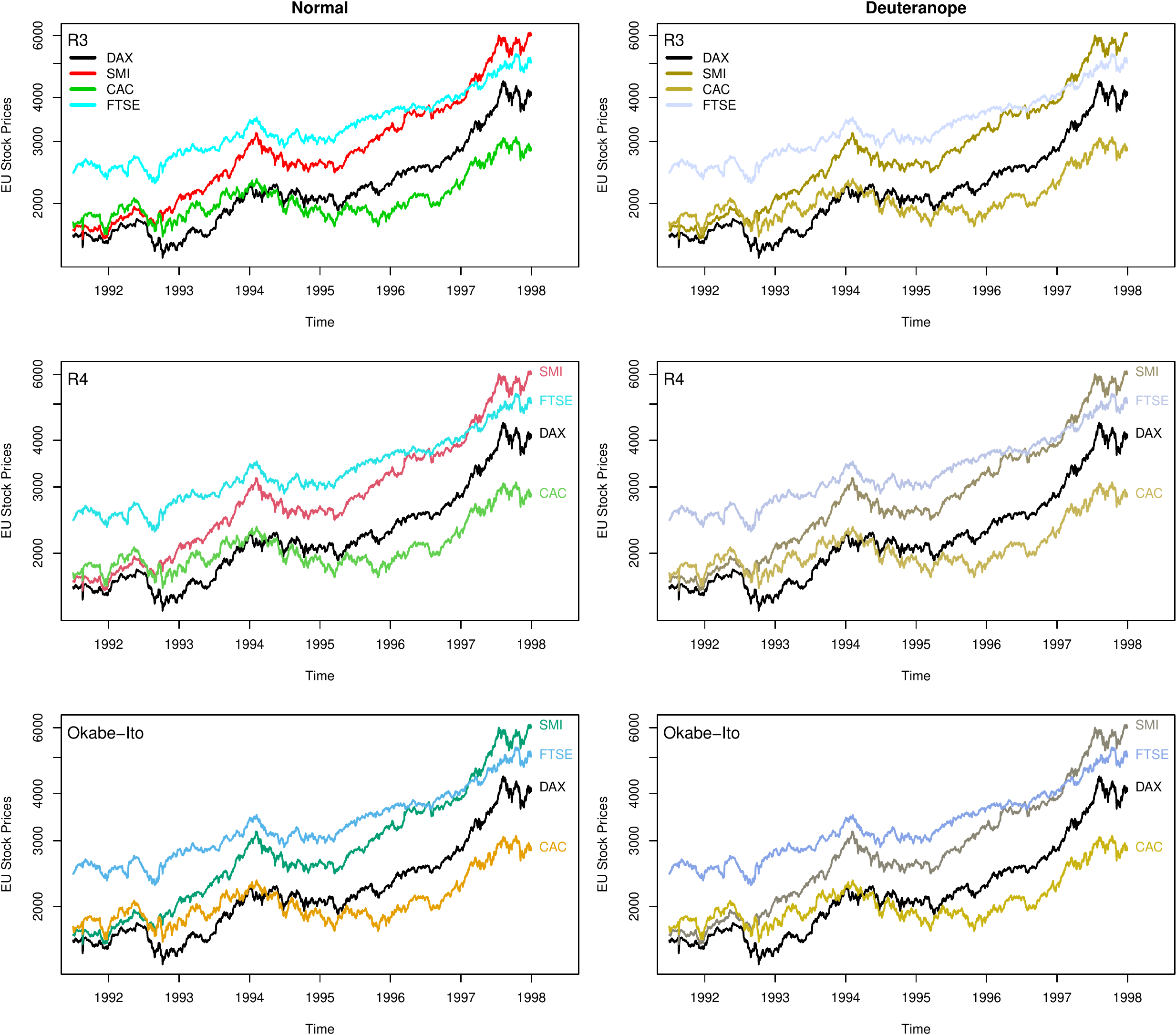} 

}

\caption{Time series line plot of \code{EuStockMarkets}. Rows: Old \code{"R3"} default palette (top), new \code{"R4"} default palette (middle), \code{"OkabeIto"} palette (bottom), designed to be robust under color vision deficiencies. Columns: Normal vision (left) and emulated deuteranope vision (right). A color legend is used in the first row and direct labels in the other rows.}\label{fig:tsplot}
\end{figure}

To illustrate the benefits of the new color palettes,
Figure~\ref{fig:tsplot} shows several versions of a time series plot,
depicting four different European stock indexes during most of the 1990s
(\texttt{EuStockMarkets} data). The plots compare
the old \texttt{"R3"} default palette with
the new \texttt{"R4"} default and the new qualitative palette
\texttt{"Okabe-Ito"}.
These can all be selected using \texttt{palette.colors()}.
The first row shows the \texttt{"R3"} default using a typical color legend in the top left corner;
the second column shows an emulation of a kind of red-green
color blindness known as deuteranopia using the \pkg{colorspace} package
(based on Machado, Oliveira, and Fernandes 2009).
The second row uses the \texttt{"R4"} palette and the third row
uses \texttt{"Okabe-Ito"}; both using direct labels for the different time series instead of
a color legend.

We can see that the \texttt{"R3"} colors are highly saturated and they vary in
luminance. For example, the cyan line is noticeably lighter than the others.
Futhermore, for deuteranope viewers, the CAC and the SMI lines are difficult to
distinguish from each other (exacerbated by the use of a color legend that makes
matching the lines to labels almost impossible). Moreover, the FTSE line is more
difficult to distinguish from the white background, compared to the other lines.

The \texttt{"R4"} palette is an improvement: the luminance is more even and
the colors are less saturated, plus the colors are more distinguishable for
deuteranope viewers (aided by the use of direct color labels instead of a legend).
The \texttt{"Okabe-Ito"} palette works even better, particularly for deuteranope viewers.

\begin{figure}[t!]

{\centering \includegraphics[width=1\linewidth]{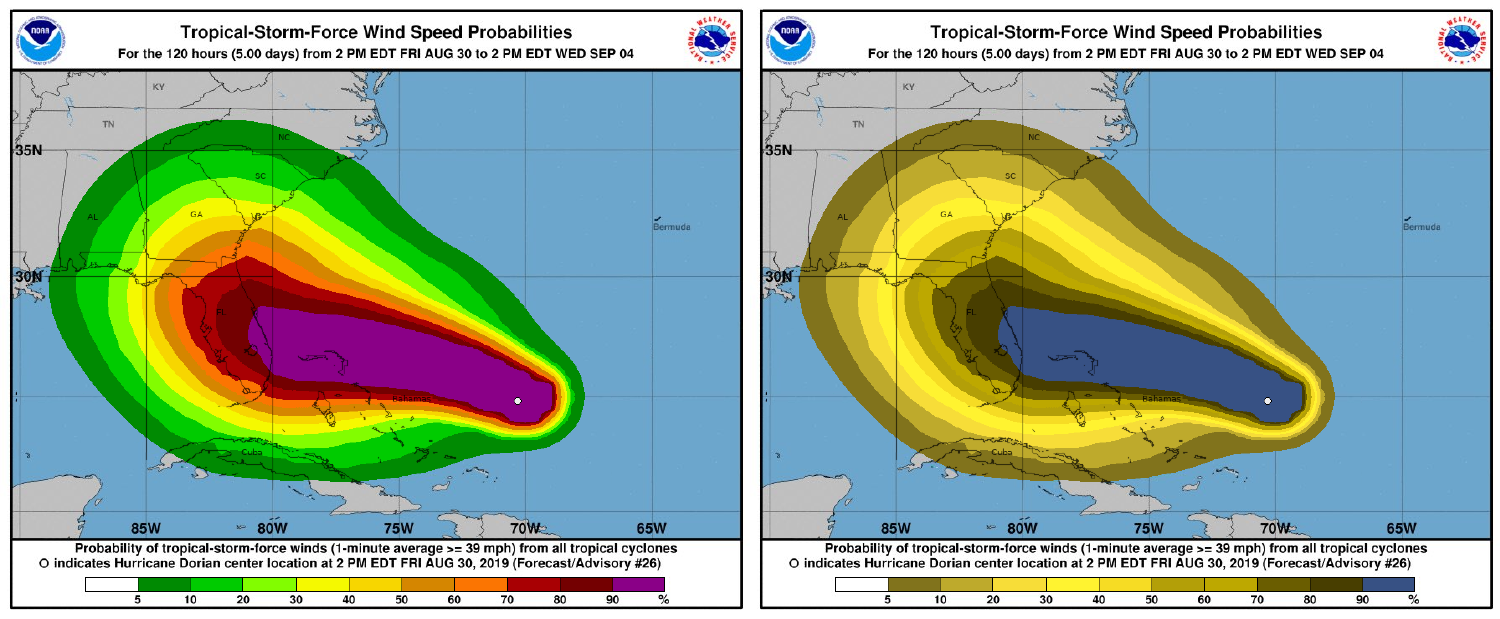} \includegraphics[width=1\linewidth]{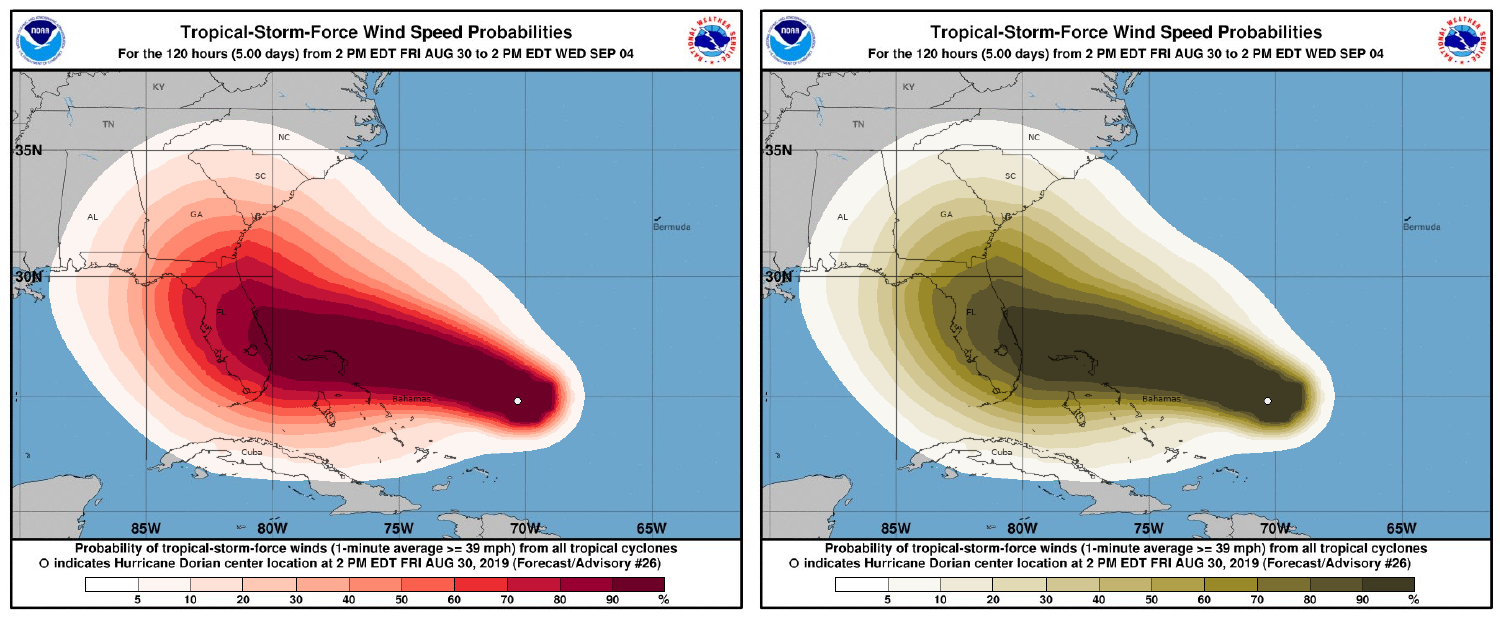} \includegraphics[width=1\linewidth]{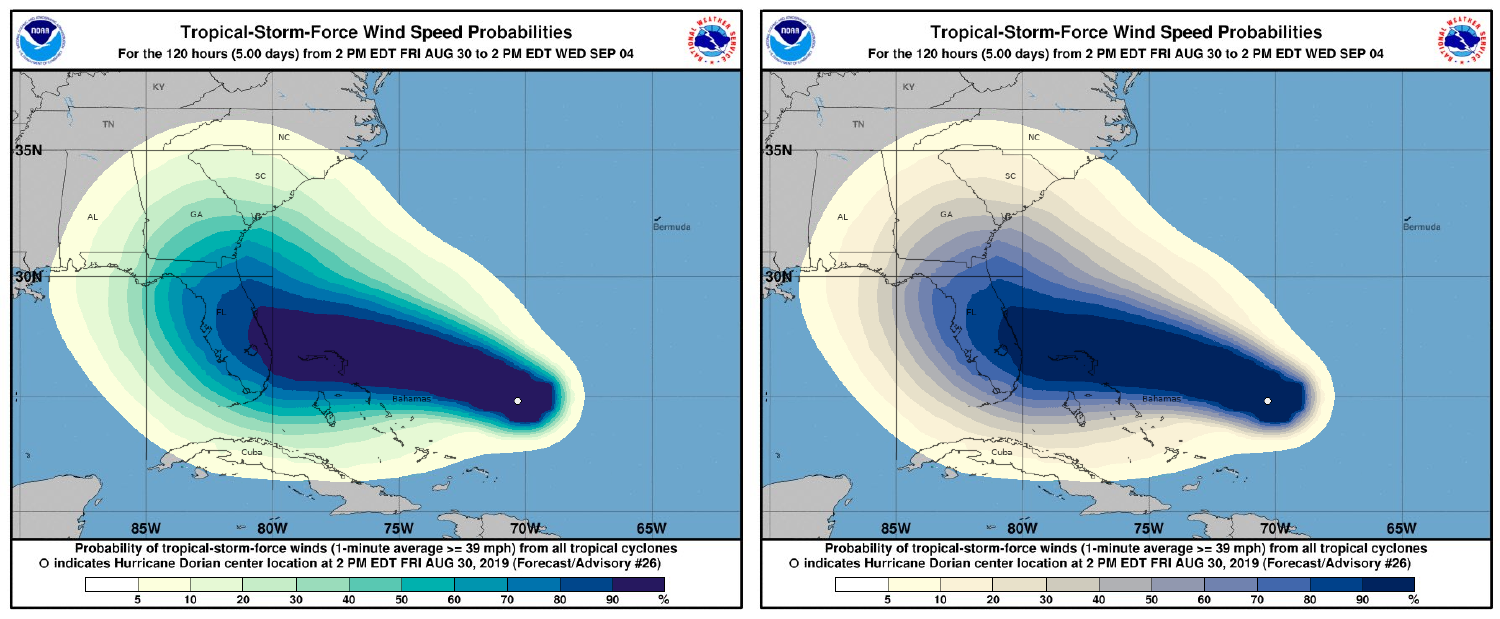} 

}

\caption{Probability of wind speeds $>$ 39\,mph (63\,km\,h$^{-1}$) during hurricane Dorian in 2019. On the left is the the original image (top row) and two reproductions using the \code{"Reds"} (middle) and \code{"YlGnBu"} (bottom) multi-hue sequential palettes. On the right are emulations of how the images on the left might appear to a colorblind viewer.}\label{fig:dorian}
\end{figure}

To illustrate an application of the new sequential color palettes
for use with continuous data,
Figure~\ref{fig:dorian} shows several versions of a weather
map that was produced by the National Oceanic and Atmospheric Administration
(and infamously misinterpreted by a former President of The United States, see Zeileis and Stauffer 2019).
The top row shows the original image along with an emulation of
deuteranopia in the second column.
The middle row uses the sequential palette
\texttt{"Reds"} that can be selected using \texttt{hcl.colors()} and the
bottom row uses the sequential palette \texttt{"YlGnBu"}, which is also
available via \texttt{hcl.colors()}.

The weather map is intended to convey
the probability of wind speeds \(>\) 39 mph
during hurricane Dorian, 2019-08-30--2019-09-04. The probabilities are highest
in the central magenta region and lowest in the outer green regions.
The original image does not convey the information very well because
there is a non-monotonic change in luminance
(from dark to light and back to dark); the high saturation across
all of the colors is also distracting. These issues persist for
deuteranope viewers, plus any benefit of a red (danger!) to green (safe)
change in hue is lost.

The \texttt{"Reds"} version of the image conveys the information more clearly
by relating the monotonic changes in probability to monotonic
changes in luminance. Hue is fairly constant in this palette and
the saturation peaks towards the middle, which is similar to the
\texttt{"Blues\ 3"} palette shown in
Figure~\ref{fig:blues3hcl}, just with a different narrow range of hues.
The deuteranope version retains this advantage.

The \texttt{"YlGnBu"} version of the image is also more effective than
the original. This palette employs a much broader range of hues and varies
chroma along with luminances so that the dark colors have higher chroma and
the light colors lower chroma
(see Figure~\ref{fig:ylgnbu-viridis}). This still clearly conveys
the order from light to dark but additionally yields more distinguishable
colors, making it easier to associate contour bands with the legend.
Note that the \texttt{"YlGnBu"} palette is similar to the very popular
\texttt{"Viridis"} palette (also shown in Figure~\ref{fig:ylgnbu-viridis}
on the right), with almost the same hue and luminance trajectories.
However, an important advantage of the \texttt{"YlGnBu"} palette in this
visualization is that the light colors have low chroma and thus signal
low risk better than the light colors in the \texttt{"Viridis"} palette which
have very high chroma. Finally, we remark that the \texttt{"YlGnBu"} version
does lose the benefit of red (danger!) at high probabilities;
an alternative would be to use the \texttt{"Purple-Yellow"} multi-hue palette
instead, a variation of which was used by Zeileis and Stauffer (2019).

The following sections describe the full range of
new color palettes in more detail. A much more condensed
overview of the new functions and palettes that are available and
some suggestions for robust default palettes are given in the
Section~\protect\hyperlink{sec:summary}{6}.

\hypertarget{a-gallery-of-palettes}{%
\section{A gallery of palettes}\label{a-gallery-of-palettes}}

This section goes through all of the color palettes that are now
available in base R (without using any additional packages).
There is some discussion of the background for the palettes,
strengths and weaknesses of different palettes, and
appropriate uses of the palettes.

\hypertarget{the-palette.colors-function-1}{%
\subsection{\texorpdfstring{The \texttt{palette.colors()} function}{The palette.colors() function}}\label{the-palette.colors-function-1}}

The \texttt{palette.colors()} function provides a range of qualitative palettes (see
Figure~\ref{fig:newPalettes} for an overview).
The first argument to the \texttt{palette.colors()} function specifies
the number of colors to return and
the \texttt{palette} argument allows us to select the palette of colors
to choose from. As previously mentioned, the default palette
is \texttt{"Okabe-Ito"}, which has very good perceptual properties.
The \texttt{"R4"} palette specifies the new R default palette
which is also returned by \texttt{palette()} by default.
As previously mentioned, this was constructed to have
reasonable perceptual properties, including accommodation for
color vision deficiencies (see Zeileis et al. 2019 for more details).
The accompanying \texttt{palette.pals()} function returns a character
vector of the available palette names.

\begin{verbatim}
palette.pals()
\end{verbatim}

\begin{verbatim}
#>  [1] "R3"              "R4"              "ggplot2"         "Okabe-Ito"      
#>  [5] "Accent"          "Dark 2"          "Paired"          "Pastel 1"       
#>  [9] "Pastel 2"        "Set 1"           "Set 2"           "Set 3"          
#> [13] "Tableau 10"      "Classic Tableau" "Polychrome 36"   "Alphabet"
\end{verbatim}

Each of the predefined palettes can be set as the default palette by passing
the palette name to the \texttt{palette()} function. For example,
the following code sets the Okabe-Ito palette as the default palette.

\begin{verbatim}
palette("Okabe-Ito")
\end{verbatim}

There are several palettes that have been taken from the color schemes of
the \pkg{ggplot2} package and Tableau (Tableau Software, LLC 2021), both
well-established graphics systems. The \texttt{"Tableau\ 10"} palette
represents the redesign of the default palette in Tableau (Stone 2016).
These palettes may be useful
to emulate the look and feel of plots from those other systems.

Several palettes come from ColorBrewer.org (Harrower and Brewer 2003) and
were originally designed for filling regions on maps. However, they
are also useful for filling regions within data visualizations like bar plots,
density plots, and heatmaps, among others.
Two of these palettes are a bit different because
they deliberately contain darker and lighter colors:
the \texttt{"Accent"} palette may be useful to emphasize one or more categories
over the others;
the \texttt{"Paired"} palette may be useful to represent more than one
categorical variable via color, e.g., different types of treatment
as well as high vs.~low levels of each treatment.

Finally, there are two palettes from the
\CRANpkg{Polychrome} package (Coombes et al. 2019).
These are much larger palettes, with colors chosen to be evenly
spread throughout HCL colorspace. The \texttt{"Polychrome\ 36"} palette
represents the largest set of colors that could be generated
while still being visually distinguishable. The \texttt{"Alphabet"}
palette is a smaller, but still large, set (one for each letter
of the alphabet). These palettes
may be useful if we are attempting to represent a very large
number of categories at once. The result is unlikely to be
easy to interpret, but these palettes will provide the best chance.

\hypertarget{the-hcl.colors-function-1}{%
\subsection{\texorpdfstring{The \texttt{hcl.colors()} function}{The hcl.colors() function}}\label{the-hcl.colors-function-1}}

The \texttt{hcl.colors()} function provides
qualitative, sequential, and diverging palettes that are derived from
certain trajectories of the perceptual properties hue, chroma, and luminance
(HCL). Most of the resulting palettes have one
or more desirable perceptual properties:

\begin{itemize}
\tightlist
\item
  \textbf{Colorblind-safe:} This means that the palette retains
  its perceptual properties for colorblind users.
\item
  \textbf{Perceptual order:} This means that there is a perceived ordering
  of the colors, typically arising from a monotonic change from
  light to dark or vice versa.
\item
  \textbf{Perceptual uniformity:} This means that if we take a small step along
  the path of the palette in HCL space, the perceived difference
  between the two colors will be the same anywhere along the path.
\item
  \textbf{Perceptual balance:} This means that, for example,
  while there are changes in hue and chroma, luminance remains
  pretty much the same, so no color stands out from the others.
\end{itemize}

These properties are very difficult to achieve in
a single palette, which is one reason why there are multiple palettes
available. Furthermore, different properties will be more or less
important depending on the data being displayed and the point that
a data visualization is attempting to make. For example, perceptual
balance is not desirable when we want to highlight a particular
point or category of interest; in that scenario we explicitly
want some colors to have a greater visual impact than others.
The details always also depend a lot on how many colors we need. For
example, a palette with light gray, medium color, and full color may still
work great on a white background if the
light gray group is less important and
is just provided in the background for
reference.

Perceptual order and colorblind-safety are closely linked
because the easiest approach to obtaining
a colorblind-safe palette is by using a monotonic
change in luminance. All of the sequential palettes in \texttt{hcl.colors()} in
fact have this property and are colorblind-safe to a certain degree, though
it depends on the luminance range how effective this is.
A quick way to check a palette for colorblind-safety is via
\texttt{colorspace::swatchplot(pal,\ cvd\ =\ TRUE)}, where \texttt{pal} is a palette of colors.
More elaborate tools
are provided by the package \CRANpkg{colorblindcheck} (Nowosad 2021).

The \CRANpkg{colorspace} package also provides functions like
\texttt{sequential\_hcl()} and \texttt{diverging\_hcl()}
to generate even further palettes by defining a custom set of
hue, chroma, and luminance
trajectories, e.g., based on specific hues that have inherent meanings
for a particular data set.

\hypertarget{qualitative-palettes}{%
\subsubsection{Qualitative palettes}\label{qualitative-palettes}}

\begin{figure}[ht!]

{\centering \includegraphics[width=1\linewidth]{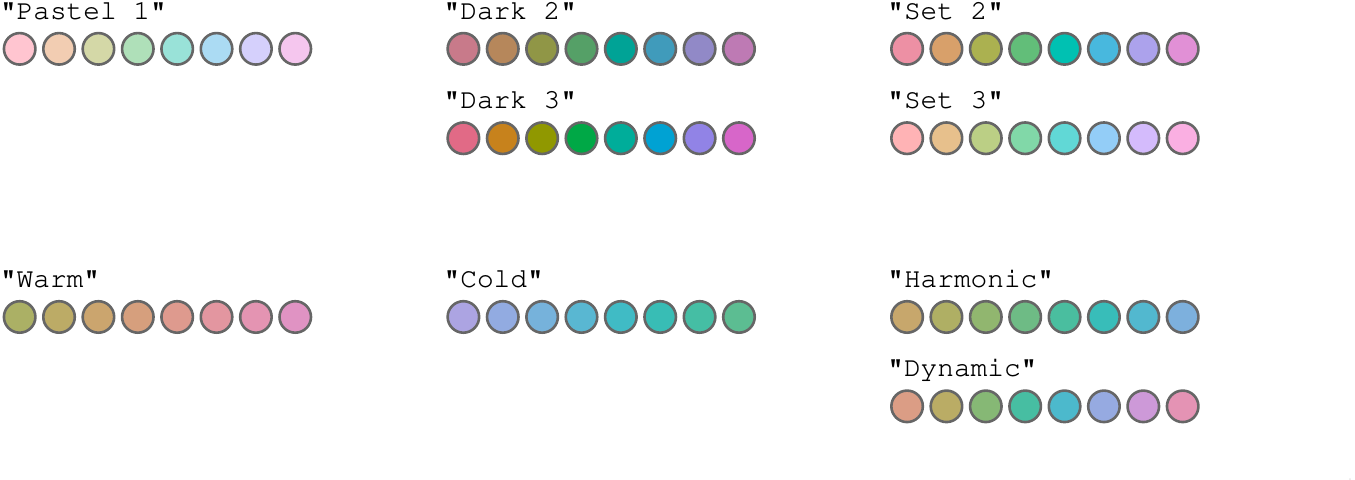} 

}

\caption{The qualitative palettes that are available with the \code{hcl.colors()} function.}\label{fig:qualPalettes}
\end{figure}

The qualitative palettes available from \texttt{hcl.colors()} are shown
in Figure~\ref{fig:qualPalettes}.
The common feature of these palettes is that they only vary hue
while using the same chroma and luminance for all of their colors.
One drawback to this approach is that fewer easily distinguishable
colors can be generated from these palettes.

The first five palettes are inspired by the ColorBrewer.org
palettes of the same name.
They employ different fixed levels of chroma and luminance and
span the full hue range.
Most of these palettes are also available as a fixed set
of colors via \texttt{palette.colors()}. There are two key differences:
First, chroma and luminance are fixed in \texttt{hcl.colors()} but typically
vary somewhat in \texttt{palette.colors()}. The former has the advantage
that the colors are more balanced. The latter has the advantage
that more sufficiently different colors can be obtained. Second,
\texttt{hcl.colors()} will return \texttt{n} colors interpolated from the full
range of hues, whereas \texttt{palette.colors()} will return the first \texttt{n}
colors from a fixed set.

The ColorBrewer.org palettes were designed with good perceptual properties
in mind, but also relied on expert opinion and trial and error.
This means that a little more care should be taken when selecting one
of the ColorBrewer-inspired HCL-based palettes because, for example,
they are often not colorblind-safe.

The remaining four palettes are taken from Ihaka (2003). These
palettes keep chroma and luminance fixed and restrict the range
of hues (blues and greens for \texttt{"Cold"} and reds and oranges for
\texttt{"Warm"}). Holding chroma and luminance fixed means that the visual impact
is even across the palette. This makes these palettes appropriate
if all categories in a variable have equal importance, but, as with
the ColorBrewer.org emulations, they are not colorblind-safe and they will not
be appropriate for grayscale printing.

When palettes are employed for shading areas in statistical displays
(e.g., in bar plots, pie charts, or regions in maps), lighter colors
(with moderate chroma and high luminance) such as \texttt{"Pastel\ 1"} or \texttt{"Set\ 3"} are typically less distracting. By contrast, when coloring points
or lines, colors with a higher chroma are often required: On
a white background a moderate luminance as in \texttt{"Dark\ 2"} or \texttt{"Dark\ 3"}
usually works better while on a black/dark background the luminance
should be higher as in \texttt{"Set\ 3"} for example.

\hypertarget{single-hue-sequential-palettes}{%
\subsubsection{Single-hue sequential palettes}\label{single-hue-sequential-palettes}}

We divide sequential palettes into single-hue (this section)
and multi-hue palettes (the next section).

Single-hue sequential palettes vary only from dark/colorful to light/gray, with
a constant underlying hue.
Figure~\ref{fig:blues3hcl} provides a good example of the hue, chroma,
and luminance trajectories for these palettes.
Certain hues will be more
appropriate for representing data on specific concepts, such
as green for ``vegetation'' and red for ``temperature''.

\begin{figure}[ht!]

{\centering \includegraphics[width=1\linewidth]{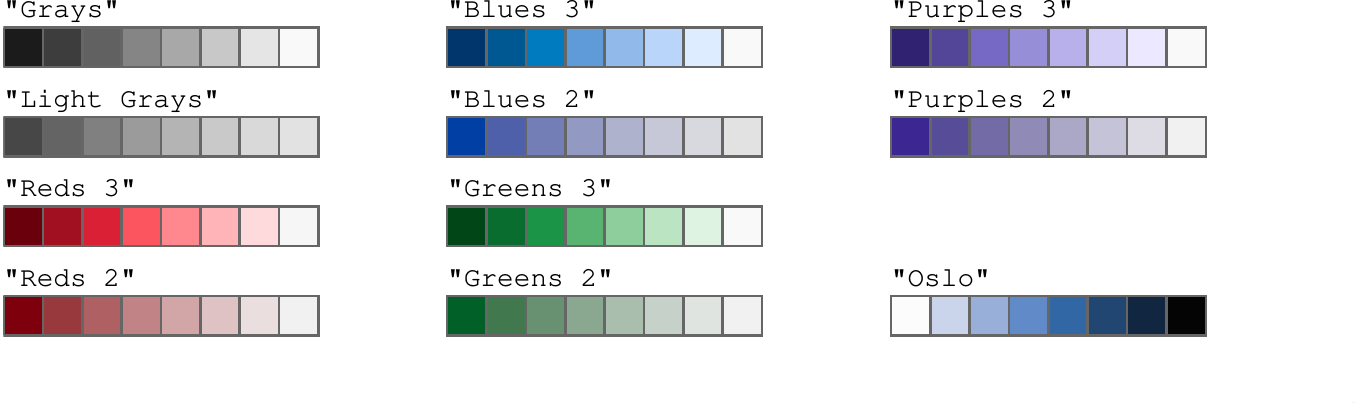} 

}

\caption{The single-hue sequential palettes that are available with the \code{hcl.colors()} function.}\label{fig:singleSeqPalettes}
\end{figure}

Figure~\ref{fig:singleSeqPalettes} shows the sequential palettes that
hold hue largely constant. All of these palettes have a large
monotonic variation in luminance, typically from dark to light. This
is also typically accompanied by a change in chroma from more
colorful to less. The result is a palette that makes it very easy to
distinguish extreme values. Some palettes also have a pronounced peak
of chroma somewhere in the middle, which makes it easier to
distinguish moderate values from extreme values (e.g.,
\texttt{"Reds\ 3"}, \texttt{"Blues\ 3"}, etc.).

All palettes in this group,
except the last one, are inspired by the ColorBrewer.org palettes with
the same base name, but are restricted to a
single hue only. They are intended for a white/light background. The
last palette, \texttt{"Oslo"}, is taken from the scientific color maps of
the \pkg{scico} package
and is intended for a black/dark background and hence the order
is reversed starting from a very light blue (almost white).

When only a few colors
are needed (e.g., for coding an ordinal categorical variable with few
levels) then a lower luminance contrast may suffice
(e.g., \texttt{"Light\ Grays"}, \texttt{"Reds\ 2"}, \texttt{"Blues\ 2"}, etc.).

\hypertarget{multi-hue-sequential-palettes}{%
\subsubsection{Multi-hue sequential palettes}\label{multi-hue-sequential-palettes}}

Multi-hue
sequential palettes not only vary luminance, from light to dark
(typically along with chroma), but also vary hue.
In order to not only bring out extreme colors in a sequential palette but also
better distinguish middle colors, it is a common strategy to employ a
sequence of hues. This leads to a large range of possible palettes.
Figure~\ref{fig:ylgnbu-viridis} shows examples of the
hue, chroma, and luminance trajectories from multi-hue palettes.

Note that the palettes in this section
differ substantially in the amount of chroma
and luminance contrasts. For example, many palettes go from a dark
high-chroma color to a neutral low-chroma color (e.g., \texttt{"Reds"},
\texttt{"Purples"}, \texttt{"Greens"}, \texttt{"Blues"}) or even light gray (e.g.,
\texttt{"Purple-Blue"}). But some palettes also employ relatively high chroma
throughout the palette (e.g., emulations of palettes from
the \pkg{viridis} and \pkg{rcartocolor} packages).
The former strategy is suitable to
emphasize extreme values,
while the latter works better if all values along the sequence should
receive the same perceptual weight.

Palettes that involve a
significant variation in hue, e.g., \texttt{"YlGnBu"}, can be more effective
when we need to match specific colors to a legend
(e.g., the bottom row of Figure~\ref{fig:dorian}) or across several
small-multiples, as in facetted plots.

\begin{figure}[ht!]

{\centering \includegraphics[width=1\linewidth]{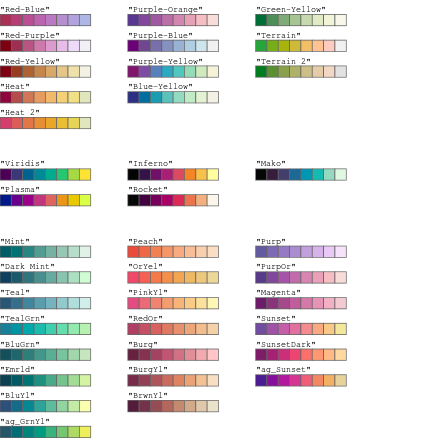} 

}

\caption{Some of the multi-hue sequential palettes that are available with the \code{hcl.colors()} function.}\label{fig:multiSeqPalettes}
\end{figure}

Of the palettes shown in Figure~\ref{fig:multiSeqPalettes}
\texttt{"Red-Blue"} to \texttt{"Terrain\ 2"} are various palettes created during the
development of the \pkg{colorspace} package.

The next collection of palettes, \texttt{"Viridis"} to \texttt{"Mako"}, emulate
popular palettes within
the Python community.
The \texttt{"Viridis"}, \texttt{"Plasma"},
and \texttt{"Inferno"} palettes come from the matplotlib Python library
and work well for identifying features of interest in false-color images.
This means that they should also work well for heatmaps.
The large range of hues means that these palettes can also serve
as qualitative palettes, which makes them robust default palettes.
However, this versatility means that
a palette that is purely sequential or purely qualitative may serve better
for a specific purpose.

The \texttt{"Mako"} and \texttt{"Rocket"} palettes are from the Seaborn Python
library with an emphasis on high chroma and a wide range of luminance.
This makes these palettes a good choice for heatmaps.

The remaining palettes in Figure~\ref{fig:multiSeqPalettes}, from
\texttt{"Mint"} to \texttt{"Sunset"} closely match palettes provided in the
\pkg{rcartocolor} package.
These palettes tend to span a much narrower range of hues, chroma,
and luminance, so can be useful if we just need to represent a
small number of ordered values.
The resulting colors from these palettes will have, for example,
more similar hues than a palette generated from \texttt{"Viridis"}, with
its wide range of hues.

\begin{figure}[ht!]

{\centering \includegraphics[width=1\linewidth]{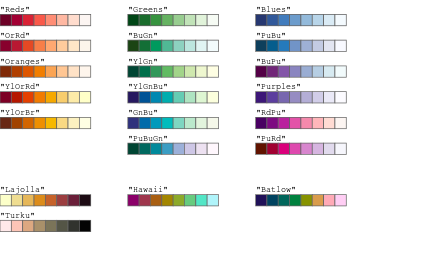} 

}

\caption{Some of the multi-hue sequential palettes that are available with the \code{hcl.colors()} function.}\label{fig:multiSeqPalettes2}
\end{figure}

Figure~\ref{fig:multiSeqPalettes2} shows the remaining multi-hue
sequential palettes that are available in \texttt{hcl.colors()}.
Most of the top group of palettes, starting with \texttt{"Reds"}, \texttt{"Greens"},
and \texttt{"Blues"}, closely match ColorBrewer.org palettes
of the same name. The \texttt{"YlGnBu"} palette is of particular note
as it uses essentially the same hues as the \texttt{"Viridis"} palette
(Figure~\ref{fig:multiSeqPalettes}), but it is more useful as a
sequential palette because chroma decreases for the high-luminance
colors (see also Figure~\ref{fig:ylgnbu-viridis}).

The next group of palettes, \texttt{"Lajolla"} to \texttt{"Batlow"} closely match the
palettes of the same name from the \pkg{scico} package.
These palettes are constructed with a luminance scale
so that there is a clear visual ordering of the palette.
They are also designed to be readable by colorblind users
and to work for grayscale printing. Finally, the palettes have been
designed for perceptual balance, so that no color has a greater
visual emphasis than any other.
Both \texttt{"Lajolla"} and \texttt{"Turku"}
are intended for use with a black/dark background.

\hypertarget{diverging-palettes}{%
\subsubsection{Diverging palettes}\label{diverging-palettes}}

\begin{figure}[ht!]

{\centering \includegraphics[width=1\linewidth]{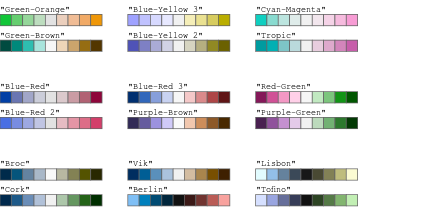} 

}

\caption{The balanced diverging palettes that are available with the \code{hcl.colors()} function.}\label{fig:diverPalettes}
\end{figure}

The diverging palettes offer a range of underlying hues
for either extreme, with either light gray or yellow as the central ``neutral''
value. The palettes with yellow at the centre provide less of a
change in colorfulness, so the ``neutral'' value is more of
a turning point rather than a local minimum.\\
Figure~\ref{fig:diverPalettes} shows the selection of diverging
palettes for use with \texttt{hcl.colors()}.

All of these palettes are ``balanced'' in the sense that chroma and luminance
vary in the same way as we move from the central neutral color towards
either end of the palette.
Figure~\ref{fig:purplegreen-fall} (left) shows this idea of balance for
the \texttt{"Purple-Green"} palette.

When choosing a particular palette for a display similar
considerations apply as for the sequential palettes. For example,
large luminance differences are important when many colors are used
while smaller luminance contrasts may suffice for palettes with fewer
colors.

Almost all of the palettes in the first two groups, those
involving simple color pairs like
\texttt{"Blue-Red"} or \texttt{"Cyan-Magenta"}, were developed as part of the
\pkg{colorspace} package, taking inspiration
from various other palettes, including more balanced and simplified
versions of several ColorBrewer.org palettes.
The exception is the \texttt{"Tropic"} palette,
which closely matches the palette of the same name from the
\pkg{rcartocolor} package.

The palettes \texttt{"Broc"} to \texttt{"Vik"} and \texttt{"Berlin"} to \texttt{"Tofino"} closely
match the scientific color maps from the \pkg{scico} package,
where the first three are intended for a white/light background and
the other three for a black/dark background.

\hypertarget{flexible-diverging-palettes}{%
\subsubsection{Flexible diverging palettes}\label{flexible-diverging-palettes}}

\begin{figure}[ht!]

{\centering \includegraphics[width=1\linewidth]{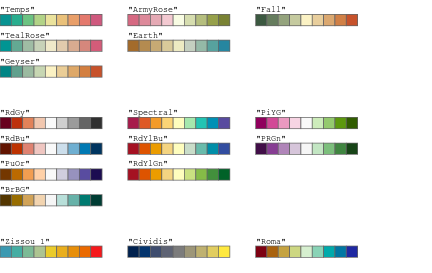} 

}

\caption{The flexible diverging palettes that are available with the \code{hcl.colors()} function.}\label{fig:diverxPalettes}
\end{figure}

Figure~\ref{fig:diverxPalettes} shows a set of
more flexible diverging
palettes. These do not impose
any restrictions that the two ``arms'' of the palette need to be
balanced and also may go through a non-gray neutral color (typically
light yellow). Consequently, the chroma/luminance within these palettes
can be rather unbalanced.
For example, Figure~\ref{fig:purplegreen-fall} (right) demonstrates this
feature of the \texttt{"Fall"} palette.

The first group of palettes, including
\texttt{"ArmyRose"} and \texttt{"Temps"} closely match the palettes of the same name
from the \pkg{rcartocolor} package.

The next group, based on two or three hues, like \texttt{"PuOr"} and \texttt{"RdYlGn"}
closely match the
palettes of the same name from ColorBrewer.org.

The final group contains
\texttt{"Zissou\ 1"}, which closely matches the palette of the same name from
the \CRANpkg{wesanderson} package (Ram and Wickham 2018), \texttt{"Cividis"},
which is an even more colorblind-safe version of \texttt{"Viridis"} (from
the \pkg{viridis} package) and
\texttt{"Roma"}, which closely matches the palette of the same name
from the \pkg{scico} package.

\hypertarget{new-defaults-in-graphical-functions}{%
\section{New defaults in graphical functions}\label{new-defaults-in-graphical-functions}}

The new default color palette will be most visible in the output from
functions in the \pkg{grDevices} and \pkg{graphics} packages.
Several functions from these packages
now have slightly different default output,
namely when they are using integer color specifications such as
\texttt{2} or \texttt{3}.
The resulting colors will still be similar to the old output, e.g.,
still a red or a green, but just a different shade.

\begin{figure}[ht!]

{\centering \includegraphics[width=1\linewidth]{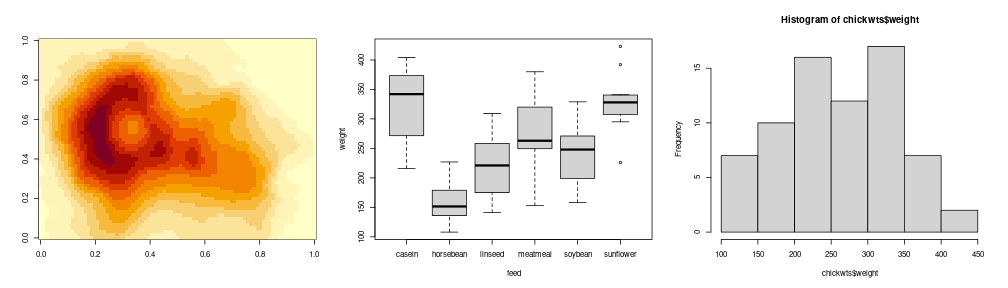} 

}

\caption{Examples of the new default color palettes that are used in the base graphics functions \code{image()}, \code{boxplot()}, and \code{hist()}.}\label{fig:graphics}
\end{figure}

Moreover, a couple of functions explicitly have new defaults:
\texttt{image()} and \texttt{filled.contour()}, now use the sequential \texttt{"YlOrRd"} palette
(from ColorBrewer)
which uses similar hues as the old \texttt{heat.colors()}.
See the left panel in Figure~\ref{fig:graphics}.

Finally, the \texttt{hist()} and \texttt{boxplot()} functions (and therefore
formula-based calls of the form \texttt{plot(num\ \textasciitilde{}\ factor,\ ...)}, also have a new default color:
light gray which makes it easier to compare the shaded areas
(see the middle and right panels in Figure~\ref{fig:graphics}).

\begin{verbatim}
image(volcano)
boxplot(weight ~ feed, data = chickwts)
hist(chickwts$weight)
\end{verbatim}

Package developers may also benefit from the new palettes available
in R; the new functions \texttt{palette.colors()} and \texttt{hcl.colors()}
allow good default palettes to be set without requiring additional
package dependencies.

\hypertarget{sec:summary}{%
\section{Summary}\label{sec:summary}}

The default color palette in R has been improved since R version 4.0.0.
The functions \texttt{palette.colors()} and \texttt{hcl.colors()}, from the
\pkg{grDevices} package, also provide a wide range of predefined
palettes based on a number of widely used graphics systems.
There are qualitative palettes for use with categorical data
and sequential and diverging palettes for use with ordinal or continuous data.
The table below summarizes the main types of palettes and provides
suggestions for good default palettes for each type.

\begin{longtable}[]{@{}
  >{\raggedright\arraybackslash}p{(\columnwidth - 6\tabcolsep) * \real{0.0791}}
  >{\raggedright\arraybackslash}p{(\columnwidth - 6\tabcolsep) * \real{0.2878}}
  >{\raggedright\arraybackslash}p{(\columnwidth - 6\tabcolsep) * \real{0.2734}}
  >{\raggedright\arraybackslash}p{(\columnwidth - 6\tabcolsep) * \real{0.3453}}@{}}
\caption{An overview of the new palette functionality: For each main type of palette, the \emph{Purpose} row describes what sort of data the type of palette is appropriate for, the \emph{Generate} row gives the functions that can be used to generate palettes of that type, the \emph{List} row names the functions that can be used to list available palettes, and the \emph{Robust} row identifies one or two good default palettes of that type.}\tabularnewline
\toprule()
\begin{minipage}[b]{\linewidth}\raggedright
\end{minipage} & \begin{minipage}[b]{\linewidth}\raggedright
Qualitative
\end{minipage} & \begin{minipage}[b]{\linewidth}\raggedright
Sequential
\end{minipage} & \begin{minipage}[b]{\linewidth}\raggedright
Diverging
\end{minipage} \\
\midrule()
\endfirsthead
\toprule()
\begin{minipage}[b]{\linewidth}\raggedright
\end{minipage} & \begin{minipage}[b]{\linewidth}\raggedright
Qualitative
\end{minipage} & \begin{minipage}[b]{\linewidth}\raggedright
Sequential
\end{minipage} & \begin{minipage}[b]{\linewidth}\raggedright
Diverging
\end{minipage} \\
\midrule()
\endhead
\emph{Purpose} & Categorical data & Ordered or numeric data
(high\(~\rightarrow~\)low) & Ordered or numeric data with a central value
(high\(~\leftarrow~\)neutral\(~\rightarrow~\)low) \\
\emph{Generate} & \texttt{palette.colors()},
\texttt{hcl.colors()} & \texttt{hcl.colors()} & \texttt{hcl.colors()} \\
\emph{List} & \texttt{palette.pals()},
\texttt{hcl.pals("qualitative")} & \texttt{hcl.pals("sequential")} & \texttt{hcl.pals("diverging")},
\texttt{hcl.pals("divergingx")} \\
\emph{Robust} & \texttt{"Okabe-Ito"}, \texttt{"R4"} & \texttt{"Blues\ 3"}, \texttt{"YlGnBu"}, \texttt{"Viridis"} & \texttt{"Purple-Green"}, \texttt{"Blue-Red\ 3"} \\
\bottomrule()
\end{longtable}

\hypertarget{references}{%
\section*{References}\label{references}}
\addcontentsline{toc}{section}{References}

\hypertarget{refs}{}
\begin{CSLReferences}{1}{0}
\leavevmode\vadjust pre{\hypertarget{ref-Bartram+Patra+Stone:2017}{}}%
Bartram, Lyn, Abhisekh Patra, and Maureen Stone. 2017. {``Affective Color in Visualization.''} In \emph{Proceedings of the 2017 CHI Conference on Human Factors in Computing Systems}, 1364--74. New York: Association for Computing Machinery. \url{https://doi.org/10.1145/3025453.3026041}.

\leavevmode\vadjust pre{\hypertarget{ref-Cleveland+McGill:1984}{}}%
Cleveland, William S., and Robert McGill. 1984. {``Graphical Perception: Theory, Experimentation, and Application to the Development of Graphical Methods.''} \emph{Journal of the American Statistical Association} 79 (387): 531--54. \url{https://doi.org/10.1080/01621459.1984.10478080}.

\leavevmode\vadjust pre{\hypertarget{ref-Polychrome}{}}%
Coombes, Kevin R., Guy Brock, Zachary B. Abrams, and Lynne V. Abruzzo. 2019. {``{Polychrome}: Creating and Assessing Qualitative Palettes with Many Colors.''} \emph{Journal of Statistical Software, Code Snippets} 90 (1): 1--23. \url{https://doi.org/10.18637/jss.v090.c01}.

\leavevmode\vadjust pre{\hypertarget{ref-Etchebehere+Fedorovskaya:2017}{}}%
Etchebehere, Sergio, and Elena Fedorovskaya. 2017. {``On the Role of Color in Visual Saliency.''} \emph{Electronic Imaging} 2017 (14): 58--63. \url{https://doi.org/10.2352/ISSN.2470-1173.2017.14.HVEI-119}.

\leavevmode\vadjust pre{\hypertarget{ref-viridis}{}}%
Garnier, Simon. 2022. \emph{{viridis}: Default Color Maps from {matplotlib}}. \url{https://CRAN.R-project.org/package=viridis}.

\leavevmode\vadjust pre{\hypertarget{ref-Harrower+Brewer:2003}{}}%
Harrower, Mark A., and Cynthia A. Brewer. 2003. {``{ColorBrewer.org}: An Online Tool for Selecting Color Schemes for Maps.''} \emph{The Cartographic Journal} 40: 27--37. \url{https://ColorBrewer.org/}.

\leavevmode\vadjust pre{\hypertarget{ref-Ihaka:2003}{}}%
Ihaka, Ross. 2003. {``Colour for Presentation Graphics.''} In \emph{Proceedings of the 3rd International Workshop on Distributed Statistical Computing, Vienna, Austria}, edited by Kurt Hornik, Friedrich Leisch, and Achim Zeileis. \url{https://www.R-project.org/conferences/DSC-2003/Proceedings/}.

\leavevmode\vadjust pre{\hypertarget{ref-Lonsdale+Lonsdale:2019}{}}%
Lonsdale, Maria Dos Santos, and David Lonsdale. 2019. {``{Design2Inform}: Information Visualisation.''} Report. The Office of the Chief Scientific Advisor, Gov UK. \url{https://eprints.whiterose.ac.uk/141931/}.

\leavevmode\vadjust pre{\hypertarget{ref-Machado+Oliveira+Fernandes:2009}{}}%
Machado, Gustavo M., Manuel M. Oliveira, and Leandro A. F. Fernandes. 2009. {``A Physiologically-Based Model for Simulation of Color Vision Deficiency.''} \emph{IEEE Transactions on Visualization and Computer Graphics} 15 (6): 1291--98. \url{https://doi.org/10.1109/TVCG.2009.113}.

\leavevmode\vadjust pre{\hypertarget{ref-RColorBrewer}{}}%
Neuwirth, Erich. 2022. \emph{{RColorBrewer}: ColorBrewer Palettes}. \url{https://CRAN.R-project.org/package=RColorBrewer}.

\pagebreak

\leavevmode\vadjust pre{\hypertarget{ref-rcartocolor}{}}%
Nowosad, Jakub. 2019. \emph{{rcartocolor}: {``CARTOColors''} Palettes}. \url{https://CRAN.R-project.org/package=rcartocolor}.

\leavevmode\vadjust pre{\hypertarget{ref-colorblindcheck}{}}%
---------. 2021. \emph{{colorblindcheck}: Check Color Palettes for Problems with Color Vision Deficiency}. \url{https://CRAN.R-project.org/package=colorblindcheck}.

\leavevmode\vadjust pre{\hypertarget{ref-Okabe+Ito:2008}{}}%
Okabe, Masataka, and Kei Ito. 2008. {``Color Universal Design ({CUD}): How to Make Figures and Presentations That Are Friendly to Colorblind People.''} \url{https://jfly.uni-koeln.de/color/}.

\leavevmode\vadjust pre{\hypertarget{ref-scico}{}}%
Pedersen, Thomas Lin, and Fabio Crameri. 2022. \emph{{scico}: Colour Palettes Based on the Scientific Colour-Maps}. \url{https://CRAN.R-project.org/package=scico}.

\leavevmode\vadjust pre{\hypertarget{ref-wesanderson}{}}%
Ram, Karthik, and Hadley Wickham. 2018. \emph{{wesanderson}: A {W}es {A}nderson Palette Generator}. \url{https://CRAN.R-project.org/package=wesanderson}.

\leavevmode\vadjust pre{\hypertarget{ref-lattice}{}}%
Sarkar, Deepayan. 2008. \emph{{lattice}: Multivariate Data Visualization with {R}}. New York: Springer-Verlag. \url{https://lmdvr.R-Forge.R-project.org/}.

\leavevmode\vadjust pre{\hypertarget{ref-Stone:2016}{}}%
Stone, Maureen. 2016. {``How We Designed the New Color Palettes in {Tableau} 10.''} \url{https://www.tableau.com/about/blog/2016/7/colors-upgrade-tableau-10-56782}.

\leavevmode\vadjust pre{\hypertarget{ref-Tableau}{}}%
Tableau Software, LLC. 2021. {``{Tableau}: Business Intelligence and Analytics Software.''} \url{https://www.tableau.com/}.

\leavevmode\vadjust pre{\hypertarget{ref-cols4all}{}}%
Tennekes, Martijn. 2023. \emph{{cols4all}: Colors for All}. \url{https://CRAN.R-project.org/package=cols4all}.

\leavevmode\vadjust pre{\hypertarget{ref-Ware:2012}{}}%
Ware, Colin. 2012. \emph{Information Visualization: Perception for Design}. 3rd ed. Morgan Kaufmann. \href{https://10.1016/C2009-0-62432-6}{10.1016/C2009-0-62432-6}.

\leavevmode\vadjust pre{\hypertarget{ref-ggplot2}{}}%
Wickham, Hadley. 2016. \emph{{ggplot2}: Elegant Graphics for Data Analysis}. 2nd ed. Springer-Verlag. \url{https://ggplot2.tidyverse.org/}.

\leavevmode\vadjust pre{\hypertarget{ref-Wiki+HCL}{}}%
Wikipedia. 2023. {``{HCL Color Space} --- {W}ikipedia{,} the Free Encyclopedia.''} URL \url{https://en.wikipedia.org/wiki/HCL_color_space}, accessed 2023-03-08.

\leavevmode\vadjust pre{\hypertarget{ref-Zeileis+Fisher+Hornik:2020}{}}%
Zeileis, Achim, Jason C. Fisher, Kurt Hornik, Ross Ihaka, Claire D. McWhite, Paul Murrell, Reto Stauffer, and Claus O. Wilke. 2020. {``{colorspace}: A Toolbox for Manipulating and Assessing Colors and Palettes.''} \emph{Journal of Statistical Software} 96 (1): 1--49. \url{https://doi.org/10.18637/jss.v096.i01}.

\leavevmode\vadjust pre{\hypertarget{ref-Zeileis+Hornik+Murrell:2009}{}}%
Zeileis, Achim, Kurt Hornik, and Paul Murrell. 2009. {``Escaping {RGB}land: Selecting Colors for Statistical Graphics.''} \emph{Computational Statistics \& Data Analysis} 53: 3259--70. \url{https://doi.org/10.1016/j.csda.2008.11.033}.

\leavevmode\vadjust pre{\hypertarget{ref-grDevices}{}}%
Zeileis, Achim, and Paul Murrell. 2019. {``{HCL}-Based Color Palettes in {grDevices}.''} \url{https://blog.R-project.org/2019/04/01/hcl-based-color-palettes-in-grdevices/}.

\leavevmode\vadjust pre{\hypertarget{ref-grDevices2}{}}%
Zeileis, Achim, Paul Murrell, Martin Maechler, and Deepayan Sarkar. 2019. {``A New \texttt{palette()} for {R}.''} \url{https://blog.R-project.org/2019/11/21/a-new-palette-for-r/}.

\leavevmode\vadjust pre{\hypertarget{ref-Zeileis+Stauffer:2019}{}}%
Zeileis, Achim, and Reto Stauffer. 2019. {``Was {T}rump Confused by Rainbow Color Map?''} \url{https://www.zeileis.org/news/dorian_rainbow/}.

\end{CSLReferences}

\bibliography{color.bib}

\address{%
Achim Zeileis\\
Universität Innsbruck\\%
Department of Statistics\\
\url{https://www.zeileis.org/}\\%
\textit{ORCiD: \href{https://orcid.org/0000-0003-0918-3766}{0000-0003-0918-3766}}\\%
\href{mailto:Achim.Zeileis@R-project.org}{\nolinkurl{Achim.Zeileis@R-project.org}}%
}

\address{%
Paul Murrell\\
University of Auckland\\%
Department of Statistics\\
\url{https://www.stat.auckland.ac.nz/~paul/}\\%
\textit{ORCiD: \href{https://orcid.org/0000-0002-3224-8858}{0000-0002-3224-8858}}\\%
\href{mailto:paul@stat.auckland.ac.nz}{\nolinkurl{paul@stat.auckland.ac.nz}}%
}

\end{article}

\end{document}